# A Cost-Effective Strategy for Storing Scientific Datasets with Multiple Service Providers in the Cloud

Dong Yuan[1], Lizhen Cui[2], Xiao Liu[3], Erjiang Fu[4], Yun Yang[5]

[1]School of Electrical and Information Engineering
The University of Sydney
Sydney, Australia
dong.yuan@sydney.edu.au

[2]School of Computer Science and Technology
Shandong University
Jinan, China
clz@sdu.edu.cn

[3]School of Information Technology
Deakin University
Melbourne, Australia
xiao.liu@deakin.edu.au

[4]Australian Bureau of Meteorology
Melbourne, Australia
ffu@bom.gov.au

[5]School of Software and Electrical Engineering
Swinburne University of Technology
Melbourne, Australia
yyang@swin.edu.au

## Abstract

Cloud computing provides scientists a platform that can deploy computation and data intensive applications without infrastructure investment. With excessive cloud resources and a decision support system, large generated datasets can be flexibly 1) stored locally in the current cloud, 2) deleted and re-generated whenever reused or 3) transferred to cheaper cloud service for storage. However, due to the pay-as-you-go model, the total application cost largely depends on the usage of computation, storage and bandwidth resources, hence cutting the cost of cloud-based data storage becomes a big concern for deploying scientific applications in the cloud. In this paper, we propose a novel strategy that can cost-effectively store large generated datasets with multiple cloud service providers. The strategy is based on a novel algorithm that finds the trade-off among computation, storage and bandwidth costs in the cloud, which are three key factors for the cost of data storage. Both general (random) simulations conducted with popular cloud service providers' pricing models and three specific case studies on real world scientific applications show that the proposed storage strategy is highly cost effective and practical for runtime utilisation in the cloud.

# 1 INTRODUCTION

With the rapid growth of e-science, domain scientists increasingly rely on computer systems to conduct their research [5] [16] [23] [26], e.g. cluster, grid and HPC (High Performance Computing) systems. In recent years, cloud computing is emerging as the latest parallel and distributed computing paradigm which provides redundant, inexpensive and scalable resources on demand to user requirements [13]. The emergence of cloud computing offers a new way for deploying scientific applications. IaaS (Infrastructure as a Service) is a very popular way to deliver services in the cloud [1], where the heterogeneity of computing systems [38] of one service provider can be well shielded by virtualisation technology. Hence, scientists can deploy their applications in unified cloud resources such as computing, storage and network services without any infrastructure investment, and only pay for their usage according to the pay-as-you-go model.

However, along with the convenience brought by using on-demand cloud services, users have to pay for the resources used, which can be substantial. Especially, nowadays scientific applications are getting more and more data intensive [11] [21] [28], where generated datasets are often gigabytes, terabytes, or even petabytes in size. As reported by Szalay et al. in [27], science is in an exponential world and the amount of application data will double every year over the next decade and future. These generated data contain important intermediate or final results of computation, which may need to be stored for reuse [7] and sharing [8]. Hence, cutting the cost of cloud-based data storage in a pay-as-you-go fashion becomes a big concern for deploying scientific applications in the cloud.

In the cloud, users have multiple options to cope with the large generated application data with a decision support system. As excessive storage and processing power can be obtained on-demand from commercial service providers, users can either store all data in the cloud and pay for the storage cost, or delete them and pay for the computation cost to regenerate them whenever they are reused. Furthermore, as cloud computing is such a fast growing market, more and more different cloud service providers with cost-effective storage solutions appear [3]. This phenomenon allows users to transfer the generated application data to cheaper services for storage with paying for the incurred bandwidth cost. Hence, in the cloud, users can flexibly store their data with different storage strategies which also lead to different total costs correspondingly. In light of this, a good storage strategy should be able to balance the usage of computation, storage and bandwidth resources in the cloud, which are three key factors for the cost of storing generated application data. Existing work [35] only investigates the trade-off between computation and storage in one cloud service provider, where bandwidth cost has not been considered.

In this paper, by investigating the trade-off among computation, storage and bandwidth, we propose a novel cost-effective runtime strategy for storing the generated application datasets in the cloud as a decision support system. We utilise a Data Dependency Graph (DDG) to represent generated application data in the cloud [35] and design the novel T-CSB algorithm which can calculate the Trade-off among Computation, Storage and Bandwidth (T-CSB) in the cloud. Based on the T-CSB algorithm, we propose a cost-effective runtime strategy for storing the generated application data with multiple service providers in the cloud.

**This paper is a significantly extended version of our conference paper [32].** The extensions are from the following aspects: 1) more comprehensive description of the T-CSB algorithm, including the new theorem, figure and algorithm pseudo code; 2) utilisation of the T-CSB algorithm to a local-optimisation based cost-effective and efficient storage strategy; 3) efficiency evaluation of the proposed strategy; 4) case studies of real scientific applications of using the proposed strategy.

The remainder of this paper is organised as follows. Section 2 presents a motivating example of scientific application and analyses the research problems. Section 3 introduces some preliminaries and data storage cost model in the cloud. Section 4 presents our novel cost-effective storage strategy in detail. Section 5 describes our experimental results for evaluation. Section 6 discusses the related work. Section 7 summarises our conclusions and points out future work.

# 2 MOTIVATING EXAMPLE AND PROBLEMS ANALYSIS

In this Section, we introduce a real world application in Structural Mechanics which generates large intermediate data with various sizes, and analyse the problems of storing them in the cloud.

## 2.1 Motivating Example

Finite Element Modelling (FEM) is an important and widely used method for impact test of objects, where classic applications are split Hopkinson pressure bar test, gas gun impact test, drop hammer test, etc. In the Faculty of Engineering and Industrial Sciences, Swinburne University of Technology, researchers of the Structural Mechanics Research Group conduct FEM simulations of Aluminium Honeycombs under dynamic out-of-plane compression to analyse the impact behaviour of the material and structure. In their research, numerical simulations of the dynamic out-of-plane compression are conducted with ANSYS/LS-DYNA software which is a powerful FEM tool for modelling non-linear mechanics of solids, fluids, gases and their interaction. The FEM application has four major steps as shown in Figure 1.

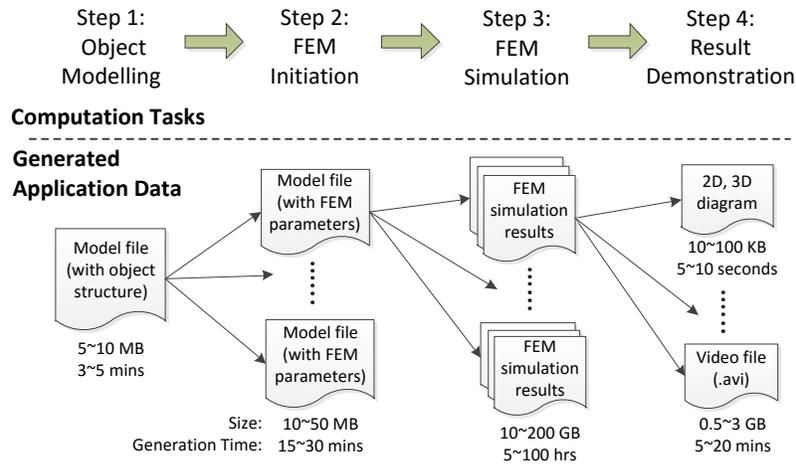

Figure 1. Overview of FEM application

From Figure 1, at beginning, based on the researchers' design, the object with special structure (i.e. the honeycombs structure in this example) for FEM analysis is generated in the Object Modelling step. Then, researchers specify more detailed parameters of the object model in the FEM Initiation step, e.g. material of the object and elements for modelling. Based on the well-defined model, researchers can run different FEM simulations according to requirements of the experiment, e.g. speed of the compression and time interval for recording data. This is the most time consuming and important step in the FEM application, which also generates the largest volume of data as simulation results. Depending on the speed of the compression, the computation time of this step varies from several hours to around one hundred hours, while depending on the time interval for recording data, the size of generated data varies from gigabytes to hundreds of gigabytes. These data are very important for researchers, based on which the simulation results can be demonstrated in various ways for analysis.

As researchers often need to run different simulations, large volume of the generated results data are accumulated as time goes on. However, due to the capacity limit of the local storage system, researchers can only store the recently generated results. Whenever they want reuse or re-analyse the results of pervious simulations, they have to re-run the simulation from beginning to regenerate the data, which is not efficient. Hence researchers consider of migrating the FEM application to the cloud where the storage bottleneck can be avoided in a cost-effective way.

## 2.2 Problem Analysis

The storage limitation would not be the case in the cloud, because the commercial cloud service providers can offer virtually unlimited storage resources. But, due to the pay-as-you-go model in the cloud, cost is one of the most important factors that users would care about. In order to make good use of the redundant cloud resources from different service providers, we need design a smart strategy to greatly reduce the cost of storing large generated application data in the cloud. However, designing this strategy is not an easy job, where the following two issues need to be carefully investigated.

1) All the resources in the cloud carry certain costs. No matter how we dealt with the generated data (e.g. storing, re-generating or transferring); we have to pay for the corresponding resources used. Different data vary in size, and have different re-generation costs and usage frequencies, e.g. data generated in the FEM application in Figure 1; therefore, it is most likely not cost effective to store all the generated data in the cloud. Intuitively, some heuristics can be applied for reducing the cost of storing the generated data. For example, we can delete the less frequently used data which have large size but small re-generation cost, and re-generate them whenever reused. Also, for the less frequently used data which have large size and huge re-generation cost, we can transfer them to cheaper places for storage, e.g. to other cloud storage systems, or even out of cloud to users' own spare storage devices. Hence, there is a trade-off among computation, storage and bandwidth in the cloud which can minimise the cost of storing the generated application data. However, finding this trade-off is not easy, as data in the cloud have dependencies (i.e. complex generation relationships) and this is the key issue in designing the cost-effective storage strategy.

2) The best trade-off among computation, storage and bandwidth may not be the best strategy for storing the generated application data. When the deleted data are needed, the regeneration not only imposes computation cost, but also causes a time delay, e.g. Step 3: FEM Simulation in Figure 1 sometimes takes several days to finish. It is also the same for data being transferred to other places are needed to be transferred back. Depending on the different time constraints of applications [20], users' tolerance of this delay may differ dramatically. Therefore, for some applications, users' preferences on storage are needed to be investigated. However, for some application, users do not concern about waiting for them to become available, hence they may delete or transfer the rarely used data to reduce the overall application cost. Therefore, this issue is not the focus of this paper.

In this paper, we **focus on the first research issue only.** We design an algorithm which can find the best trade-off among computation, storage and bandwidth in the cloud, based on which we develop a cost-effective strategy for storing generated application data with multiply cloud service providers. With regard to the **second research issue**, in our prior work [36], we have proposed an approach to incorporate users' preferences on storage by specifying corresponding parameters in the algorithm. This approach can be directly utilised in this work; hence we do not give detailed description in this paper.

# 3 SCIENTIFIC DATASETS STORAGE IN CLOUDS

In this section, we first present some preliminaries including a classification of application data in the cloud and the important concept of DDG (Data Dependency Graph). Then we present the data storage cost model which represents the trade-off among computation, storage and bandwidth in the cloud.

## 3.1 Application Data and DDG

In general, there are two types of data stored in the cloud, original data and generated data.

1) *Original data* are the data uploaded by users, for example, in scientific applications they are usually the raw data collected from the devices in the experiments. For these data, users need to decide whether they should be stored or deleted since they cannot be regenerated by the system once deleted. As cost of storing *original data* is fixed, they are **not** considered in the scope of this paper.

2) *Generated data* are the data newly produced in the cloud while the applications run. They are the intermediate or final computation results of the applications, which can be reused in the future. For these data, their storage can be decided by the system since they can be regenerated if their provenance is known. Hence, our storage strategy is **only** applied to the *generated data* in the cloud that can automatically decide the storage status of generated datasets in applications. In this paper, we refer *generated data* as **dataset(s)**.

DDG (Data Dependency Graph) [35] is a directed acyclic graph (DAG) which is based on data provenance in scientific applications. All the datasets once generated in the cloud, whether stored or deleted, their references are recorded in DDG. In other words, it depicts the generation relationships of datasets, with which the deleted datasets can be regenerated from their nearest existing preceding datasets. Figure 2 depicts a simple DDG, where every node in the graph denotes a dataset. We denote dataset $d_i$ in DDG as $d_i \in DDG$. Furthermore, $d_1$ pointing to $d_2$ means that $d_1$ is used to generate $d_2$; $d_2$ pointing to $d_3$ and $d_5$ means that $d_2$ is used to generate $d_3$ and $d_5$ based on different operations; $d_4$ and $d_6$ pointing to $d_7$ means that $d_4$ and $d_6$ are used together to generate $d_7$.

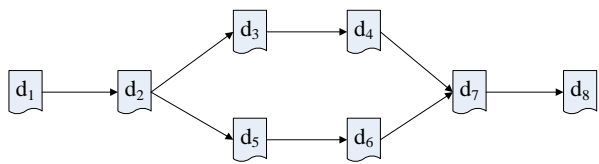

Figure 2. A simple Data Dependency Graph (DDG)

To better describe the relationships of datasets in DDG, we define a symbol: →, which denotes that two datasets have a generation relationship, where $d_i \rightarrow d_j$ means that $d_i$ is a predecessor dataset of $d_j$ in DDG. For example, in Figure 2's DDG, we have $d_1 \rightarrow d_2$, $d_1 \rightarrow d_4$, $d_5 \rightarrow d_7$, $d_1 \rightarrow d_7$, etc. Furthermore, → is transitive, i.e.

$$d_i \rightarrow d_j \rightarrow d_k \Leftrightarrow d_i \rightarrow d_j \wedge d_j \rightarrow d_k \Rightarrow d_i \rightarrow d_k.$$

## 3.2 Datasets Storage Cost Model

In a commercial cloud computing environment, service providers have their cost models to charge users. In general, there are three basic types of resources in the cloud: computation, storage and bandwidth. Popular cloud services providers' cost models are based on these types of resources. For example, Amazon cloud services' prices are as follows[2]:

$0.10 per CPU instance hour for the computation resources;
$0.15 per Gigabyte per month for the storage resources;
$0.12 per Gigabyte bandwidth resources for data downloaded from Amazon via Internet.

In this paper, we facilitate our datasets storage cost model in the cloud as follows:

***Cost = Computation + Storage + Bandwidth***

where the total cost of the datasets storage, *Cost*, is the sum of *Computation*, which is the total cost of computation resources used to regenerate datasets, *Storage*, which is the total cost of storage resources used to store the datasets, and *Bandwidth*, which is the total cost of bandwidth resources used for transferring datasets.

---

[2] The prices may fluctuate from time to time according to market factors. As this paper's focus is on cost effectiveness, to simplify the problem, we assume that the same types of computation resources are used for generatio and regeneration of datasets, and the same types of storage resources are used for storing datasets.

To utilise the datasets storage cost model, we assume that the application be deployed in one cloud service[3], denoted as $c_1$, and there be $m$ different cloud services, denoted as $\{c_1, c_2, ... c_m\}$, for storing the generated datasets in the cloud. For a dataset in DDG $\{d_1, d_2, ... d_n\}$, denoted as $d_i \in DDG$, we define its attributes as follows: $<x_i, y_{i,s}, z_{i,s}, f_i, v_i, provSet_i, CostR_i>$[4], where

- $x_i$ denotes the generation cost of dataset $d_i$ from its direct predecessors in the cloud.
- $y_{i,s}$ denotes the cost per time unit (i.e. storage cost rate) of storing dataset $d_i$ in cloud service $c_s$. Especially, $y_{i,1}$ denotes the cost rate of storing $d_i$ in the cloud service where the application is deployed.
- $z_{i,s}$ denotes the transfer cost of dataset $d_i$ from service provider $c_s$ to $c_1$, especially, $z_{i,1} = 0$.
- $f_i$ is a flag which denotes the storage status of dataset $d_i$. Specifically, $f_i = s$, $s \in \{1,2,...m\}$ represents that dataset $d_i$ is stored in cloud service $c_s$, and $f_i = 0$ represents that dataset $d_i$ is deleted.
- $v_i$ denotes the usage frequency, which indicates how often $d_i$ is used.
- $provSet_i$ denotes the set of stored provenance that are needed when regenerating dataset $d_i$. If we want to regenerate $d_i$, we have to find its direct predecessors, which may also be deleted or stored in other cloud services. $provSet_i$ is the set of the nearest stored predecessors of $d_i$ in the DDG. Hence the generation cost of $d_i$ is

$$genCost(d_i) = \sum_{\{j|d_j \in provSet_i\}} z_{j,s} \\ + \sum_{\{k|d_j \in provSet_i \wedge d_j \to d_k \to d_i\}} x_k + x_i \quad (1)$$

As we can see from formula (1), the regeneration cost of $d_i$ is two folds: 1) the bandwidth cost of transferring $d_i$'s stored provenance datasets to $c_1$ which is the cloud service that the application is deployed, and 2) the computation cost of regenerating $d_i$ in $c_1$.

- $CostR_i$ is $d_i$'s cost rate, which means the average cost per time unit of dataset $d_i$ in the cloud. The value of $CostR_i$ depends on the storage status of $d_i$, where

$$CostR_i = \begin{cases} genCost(d_i) * v_i, & f_i = 0 \quad //d_i \text{ is deleted} \\ z_{i,s} * v_i + y_{i,s}, & f_i = s \quad //d_i \text{ is stored in } c_s \end{cases} \quad (2)$$

Hence, the total cost rate of storing a DDG is the sum of $CostR$ of all the datasets in it, which is $\sum_{d_i \in DDG} CostR_i$. We further define the storage strategy of a DDG as $F$, which denotes the storage status of datasets in the DDG. Formally, $F = \{f_i | d_i \in DDG\}$, which is the set of every dataset's attribute $f_i$ indicating the cloud service in which $d_i$ is stored. We denote the cost rate of storing a DDG with the storage strategy $F$ as $SCR$ (Sum of Cost Rate), where

$$SCR = \left(\sum_{d_i \in DDG} CostR_i\right)_F \quad (3)$$

Based on the definition above, different storage strategies lead to different cost rates for the application. This cost rate, i.e. cost per time unit, represents the cost effectiveness of storage strategies, which incorporates the trade-off among computation, storage and bandwidth costs in the cloud. In next section, we will present the design of our cost-effective storage strategy based on this trade-off model.

---

[3] We assume that the application only run with one cloud service due to the following two reasons: 1) Some applications contain dedicate commercial software, e.g. the ANSYS/LS-DYNA software in the FEM application introduced in Section 2. Due to the license restriction, these kinds of software cannot be freely installed in different service providers' resources in the cloud. 2) Migrating applications, especially scientific applications to a cloud service is a complex process. In order to take advantage of on-demand cloud services, software in the applications usually need second development to facilitate the dynamic scale up and down in the cloud.

[4] These atrributes were introduce in our prior work [35]    D. Yuan, Y. Yang, X. Liu, and J. Chen, "On-demand Minimum Cost Benchmarking for Intermediate Datasets Storage in Scientific Cloud Workflow Systems," *Journal of Parallel and Distributed Computing*, vol. 71, pp. 316-332, 2011., based on which we incorporate bandwidth cost of data transfer into the original definitions. If needed, please refer to our prior work [35]    ibid. for more detailed description of these attributes.

# 4 COST-EFFECTIVE DATASETS STORAGE STRATEGY IN MULTIPLE CLOUD SERVICES

In the section, we first briefly introduce the philosophy of the novel T-CSB algorithm, followed by the detailed steps of the algorithm in order to find the best trade-off among computation, storage and bandwidth costs for storing datasets of linear DDG; then we introduce our cost-effective strategy for storing generated datasets with multiple cloud services in detail.

## 4.1 Overview of T-CSB (Trade-off among Computation, Storage and Bandwidth) Algorithm

In this paper, we design the T-CSB algorithm that can find the minimum cost storage strategy for storing datasets of linear DDG with multiple cloud storage services. Linear DDG means a DDG with no branches, where each dataset in the DDG only has one direct predecessor and successor except the first and last datasets. The minimum cost storage strategy found by the algorithm represents the best trade-off among computation, storage and bandwidth costs in the cloud.

The basic idea of the T-CSB algorithm is to construct a Cost Transitive Graph (CTG) based on the linear DDG. First, for every dataset in the DDG, we create a set of vertices in the CTG representing different storage services where the dataset can be stored. Next, we design smart rules for adding edges to the CTG and setting weights to them. Based on rules, we guarantee that in the CTG, the paths from the start vertex to the end vertex have a one-to-one mapping to the storage strategies of the DDG, and the length of every path equals to the cost rate of the corresponding storage strategy in the cloud. Then we can use the well-known Dijkstra shortest path algorithm (or Dijkstra algorithm for short) to find the shortest path in the CTG, which in fact represents both the minimum cost storage strategy for datasets of the DDG with multiple storage services, and the best trade-off among computation, storage and bandwidth costs in the cloud.

## 4.2 Detailed Steps in the T-CSB Algorithm

Given a linear DDG with datasets $\{d_1, d_2 \ldots d_n\}$ and $m$ cloud services $\{c_1, c_2 \ldots c_m\}$ for storage. The T-CSB algorithm has the following four steps:

**Step 1**: Create vertices for the CTG. First, we create the start and end vertices, denoted as $ver_{start}$ and $ver_{end}$. Then, for every $d_i \in DDG$, we create a vertex set $V_i = \{ver_{i,1}, ver_{i,2} \ldots ver_{i,m}\}$, where $m$ is the number of cloud services in which $d_i$ can be stored. Hence $ver_{i,s}$ represents dataset $d_i$ storing in cloud service $c_s$.

**Step 2**: Add directed edges to the CTG. For every $ver_{i,s} \in CTG$, we add out-edges to all vertices in the set of $\{ver_{i',s'} | ver_{i',s'} \in CTG \wedge d_i, d_{i'} \in DDG \wedge d_i \rightarrow d_{i'}\}$. In other words, for any two vertices $ver_{i,s}, ver_{i',s'} \in CTG$ belonging to different datasets' vertex sets (i.e. $V_i \neq V_{i'}$), we create an edge between them. Formally,

$$ver_{i,s}, ver_{i',s'} \in CTG \wedge d_i, d_{i'} \in DDG \wedge d_i \rightarrow d_{i'} \Rightarrow \exists e < ver_{i,s}, ver_{i',s'} >.$$

Especially, for $ver_{start}$, we add out-edges to all other vertices in the CTG, and for $ver_{end}$, we add in-edges from all other vertices in the CTG.

In Figure 3, we demonstrate the process of constructing CTG for a DDG with $n$ datasets and $m$ different cloud services for the storage.

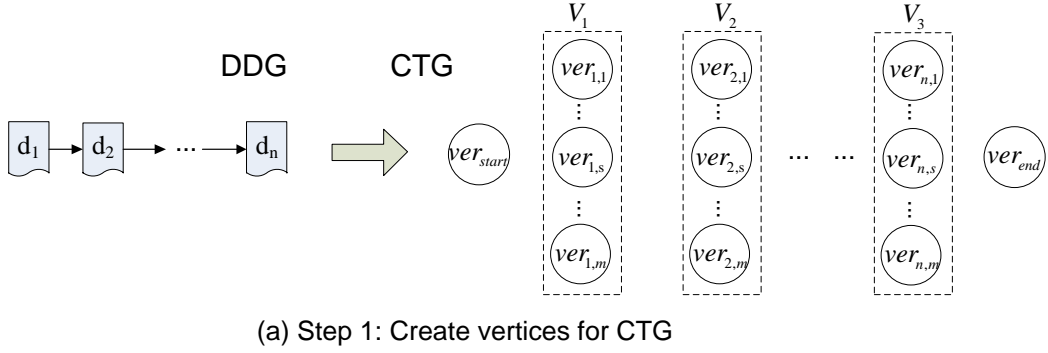

(a) Step 1: Create vertices for CTG

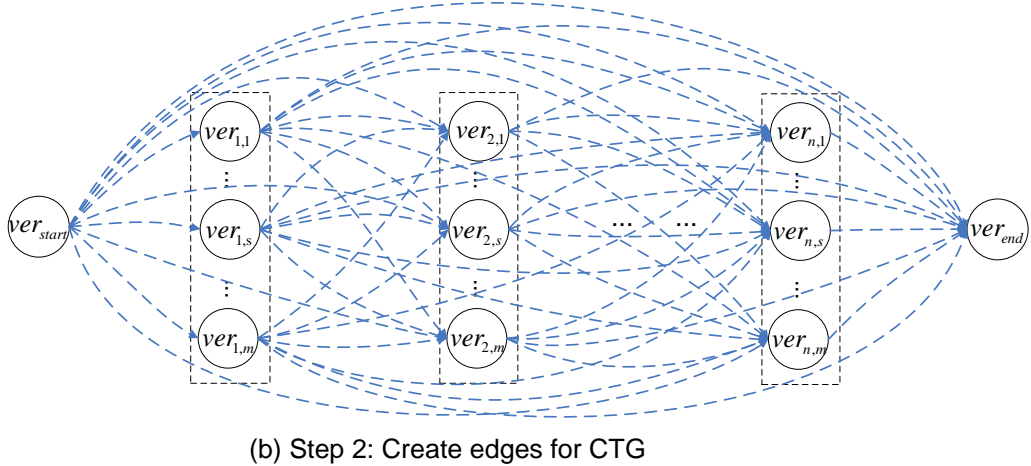

(b) Step 2: Create edges for CTG

Figure 3. Example of constructing CTG for DDG

**Step 3**: Set weights to edges in the CTG. The reason we call the graph Cost Transitive Graph is because the weights of its edges are composed of the cost rates of datasets. For an edge $e<ver_{i,s}, ver_{i',s'}>$, we denote its weight as $\omega<ver_{i,s}, ver_{i',s'}>$, which is defined as the sum of cost rates of $d_{i'}$ and the datasets between $d_i$ and $d_{i'}$, supposing that only $d_i$ and $d_{i'}$ are stored with corresponding cloud services and the rest of datasets between $d_i$ and $d_{i'}$ are all deleted. Formally:

$$\omega<ver_{i,s}, ver_{i',s'}>$$
$$= CostR_{i'} + \sum_{\{d_k | d_k \in DDG \wedge d_i \to d_k \to d_{i'}\}} CostR_k \qquad (4)$$
$$= (z_{i',s'} * v_{i'} + y_{i',s'}) + \sum_{\{d_k | d_k \in DDG \wedge d_i \to d_k \to d_{i'}\}} (genCost(d_k) * v_k)$$

Since we are discussing linear DDG, for the datasets between $d_i$ and $d_{i'}$, $d_i$ is the only dataset in their *provSet*s. Hence we can further derive:

$$\omega<ver_{i,s}, ver_{i',s'}> = (z_{i',s'} * v_{i'} + y_{i',s'})$$
$$+ \sum_{\{d_k | d_k \in DDG \wedge d_i \to d_k \to d_{i'}\}} \left( \left( z_{i,s} + x_k + \sum_{\{d_h | d_h \in DDG \wedge d_i \to d_h \to d_k\}} x_h \right) * v_k \right)$$

**Step 4**: Find the shortest path of the CTG. From the above construction steps, we can clearly see that the CTG is an acyclic oriented graph. Hence we can use the Dijkstra algorithm to find the shortest path from $ver_{start}$ to $ver_{end}$. The Dijkstra algorithm is a classic greedy algorithm to find the shortest path in graph theory. We denote the shortest path from $ver_{start}$ to $ver_{end}$ as $P_{\min}<ver_{start}, ver_{end}>$.

Based on above steps of the T-CSB algorithm, we can draw the following theorem.

**Theorem**: Given a linear DDG with datasets $\{d_1, d_2 \ldots d_n\}$ and $m$ cloud services $\{c_1, c_2 \ldots c_m\}$ for storage, the length of $P_{\min} <ver_{start}, ver_{end}>$ of its CTG is the minimum cost rate for storing the datasets in the DDG, and the corresponding storage strategy is represented by the vertices that $P_{\min} <ver_{start}, ver_{end}>$ traverses.

**Proof:**

1) There is a one-to-one mapping between storage strategies of datasets in the DDG and paths from $ver_{start}$ to $ver_{end}$ in the CTG. Given any storage strategy, we can find an order of these stored datasets, since the DDG is linear. Then we can find the exact path in the CTG that has traversed the vertices representing the storage status of these datasets, e.g. $ver_{i,s} \in CTG$ represents dataset $d_i$ stored in cloud service $c_s$. Similarly, given any path from $ver_{start}$ to $ver_{end}$ in the CTG, we can find the vertices traversed, which represent a storage strategy.

2) Based on the setting of weights to the edges, the length of a path from $ver_{start}$ to $ver_{end}$ in the CTG equals to the total cost rate of the corresponding storage strategy.

3) $P_{\min} <ver_{start}, ver_{end}>$ is the shortest path from $ver_{start}$ to $ver_{end}$ as found by the Dijkstra algorithm.

**Theorem holds.**

According to this theorem, the T-CSB algorithm finds the minimum cost storage strategy for a linear DDG. The pseudo code of the T-CSB algorithm is shown in Figure 4.

**Algorithm: T-CSB**
**Input:** A linear DDG $\{d_1, d_2 \ldots d_n\}$;
Cloud service providers $\{c_1, c_2 \ldots c_m\}$;
**Output:** The minimum cost storage strategy

```
01. Create ver_start, ver_end;
02. for ( every dataset d_i in DDG )                   //Create vertices for CTG
03.     Create V_i = {ver_{i,1}, ver_{i,2}, ..., ver_{i,m}};
04. Add vertices V_1 ∪ V_2 ... ∪ V_3 ∪ {ver_start, ver_end} to CTG;
05. for ( every ver_{i,s} ∈ CTG )                      //Create edges for CTG
06.     for ( every ver_{i',s'} ∈ CTG ∧ (d_i, d_{i'} ∈ DDG ∧ d_i → d_{i'}) )
07.         Create e < ver_{i,s}, ver_{i',s'} >;       //Create an edge
08.         weight=0;                                   //Calculate the edge weight
09.         for ( every d_k ∈ DDG ∧ (d_i → d_k → d_{i'}) )
10.             genCost=0;
11.             for ( every d_h ∈ DDG ∧ (d_i → d_h → d_k) )
12.                 genCost = genCost + x_h;
13.             weight = weight + (z_{i,s} + x_k + genCost) * v_k;
14.         weight = weight + z_{i',s'} * v_{i'} + y_{i',s'};
15.         Set e < ver_{i,s}, ver_{i',s'} > = weight;  //Set edge weight
16. P = Dijkstra ( ver_start, ver_end, CTG );          //Find the shortest path
17. Return P;                                           //The minimum cost storage strategy
```

Figure 4. Pseudo code of T-CSB algorithm

## *4.3 Cost-Effective Storage Strategy*

In our prior work [36], we developed a local-optimisation based strategy for storing datasets in one cloud service, which is highly cost effective and practical. The same philosophy can be adapted in this work to derive our cost-effective storage strategy with multiple cloud services. The strategy goes as follows:

(1) Given a general DDG, we first partition it into linear segments and apply the T-CSB algorithm to calculate the storage strategy.

We search for the datasets that have multiple direct predecessors or successors (i.e. the join and split datasets in the DDG), and use these datasets as the partitioning points to divide it into linear DDG segments, as shown in Figure 5. Based on the linear DDG segments, we use the T-CSB algorithm to find their storage strategies. This is the essence for achieving cost effectiveness.

(2) When new datasets are generated in the system, they are treated as a new DDG segment and added to the old DDG. Correspondingly, its storage status is calculated in the same way as the old DDG.

(3) When a dataset's usage frequency is changed, the storage status of the linear DDG segment that contains this dataset is re-calculated.

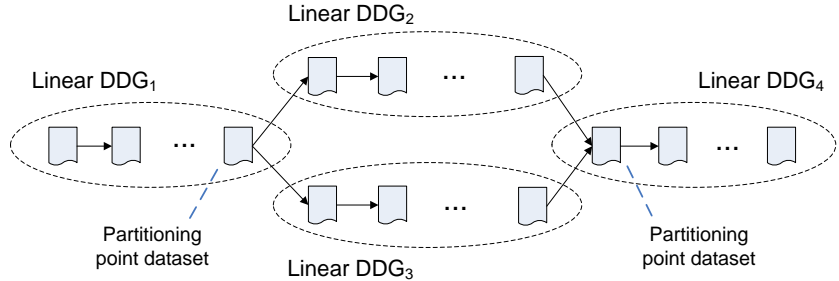

Figure 5. Dividing a DDG into linear DDG segments

By utilising the T-CSB algorithm, our strategy achieves the local-optimisation of storing datasets in the DDG. The theoretical analysis of cost effectiveness for the local-optimisation based strategy is given in our prior work [36]. In Section 5, we will demonstrate experimental results to further evaluate the cost effectiveness of our strategy.

Next, we analyse the efficiency of our storage strategy. As introduced in Section 4.2, for a linear DDG with $n$ datasets and $m$ cloud services for storage, we need to create $mn$ vertices in the CTG. Hence the number of edges in the CTG is in the magnitude of $m^2n^2$. Since the time complexity of calculating the longest edge's weight is $O(n^2)$, the worst case time complexity of the T-CSB algorithm is $O(m^2n^4)$. In our strategy, by dividing the general DDG into linear DDG segments, the time complexity is well controlled within $O(lm^2n_i^4)$, where $l$ is the number of the linear DDG segments and $n_i$ is the number of datasets in the linear DDG segments which is usually very small. In Section 5, we will further evaluate the efficiency of our strategy by experimental results.

## 5 EVALUATION

The datasets storage strategy proposed in this paper is generic. It can be used in any applications with different cloud services. In this section, we demonstrate simulation results conducted on Amazon cloud. First, we introduce our simulation setup and evaluation method. Then, we present general random simulations and evaluate the overall performance of our strategy. Next, we present three case studies about utilising our strategy in scientific applications, and use the real world data to demonstrate how our strategy works in storing the generated datasets.

## 5.1 Simulation Setup and Evaluation Method

As Amazon is a well-known and widely recognised cloud service provider, we conduct experiments on Amazon cloud using on-demand services for simulation. We implement our strategy in Java programming language and run the strategy on the virtualised EC2 instance with the Amazon Linux Image to evaluate its cost effectiveness and efficiency. We choose the standard small instance (m1.small) to conduct the experiments, because it is the basic type of EC2 CPU instances, which has a stable performance of one ECU[5].

To evaluate the cost effectiveness of our storage strategy for multiple cloud services, we compare it with different representative storage strategies for one cloud service provider, which are as follows:

- Store all datasets strategy, in which all generated datasets of the application are stored in the cloud.
- Store none datasets strategy, in which all generated datasets of the application are deleted after being used.
- Cost rate based strategy reported in [33] [37], in which we store datasets in the cloud by comparing their own generation cost rate and storage cost rate.
- Local-optimisation based strategy reported in [34] [36], in which we only achieve the localised optimum of the trade-off between computation and storage in the cloud.

Next, we assume that the scientific application be deployed in Amazon cloud using EC2 service[6] ($0.1 per CPU instance hour) for computation and S3 service ($0.15 per gigabyte per month) for storage. To utilise our storage strategy, we assume that generated datasets can be transferred to another two cloud services for storage with the prices:

- Storage Service One: $0.1 per gigabyte per month for storage and $0.01 per gigabyte for outbound[7] data transfer.
- Storage Service Two: $0.05 per gigabyte per month for storage and $0.06 per gigabyte for outbound data transfer.

We only use the above prices as representatives, as many cloud service providers (e.g. GoGrid[8], Rackspace[9], Haylix[10], and Amazon Glacier[11] etc.) have similar pricing models.

To further demonstrate the practicality of our storage strategy, we adapt real cloud service providers' pricing models and use them as the additional cloud storage service respectively in the simulation. Specifically,

(1) Amazon Glacier. Glacier is an extremely low-cost storage service that provides secure and durable storage for data archiving and backup. The pricing model for using Glacier is: $0.01 per gigabyte per month for storage, $0.02 per gigabyte for outbound data transfer from Glacier.

(2) Haylix cloud storage. Haylix is a leading Australian IaaS cloud service provider, who provides reliable cloud storage with fast access for local Australian users. As data transfer over the Internet is often expensive and relatively slow in general, some cloud service providers (e.g. Amazon) cooperate with network Infrastructure providers (e.g. Equinix) to provider dedicate connection service (e.g. AWS Direct Connect) for boosting the data transfer speed in and out of the cloud. Hence, we use the pricing models of Haylix and AWS Direct Connect in our simulation, i.e. $0.12 per gigabyte per month for storage in Haylix, $0.046 per gigabyte for outbound data transfer from Haylix.

---

[5] ECU (EC2 Computing Unit) is the basic unit defined by Amazon to measure the compute resources. Please refer to the following address for details. http://aws.amazon.com/ec2/instance-types/
[6] Amazon cloud service offers different CPU instances with different prices, where using expensive CPU instances with higher performance would reduce computation time. There exists a trade-off of time and cost [14]    S. K. Garg, R. Buyya, and H. J. Siegel, "Time and Cost Trade-Off Management for Scheduling Parallel Applications on Utility Grids," *Future Generation Computer Systems,* vol. 26, pp. 1344-1355, 2010. which is different to the trade-off of computation and storage described in this paper, hence is out of this paper's scope.
[7] At present, most cloud storage services only charge on the outbound data transfer, while inbound data transfer is usually free.
[8] GoGrid: http://www.gogrid.com/
[9] Rackspace: http://www.rackspace.com/
[10] Haylix: http://www.haylix.com/
[11] Amazon Glacier: http://aws.amazon.com/glacier/

In this section, we demonstrate some representative results. More detailed experimental results and the program code are provided at http://www.ict.swin.edu.au/personal/dyuan/doc/fgcs.zip for readers to download.

## 5.2 General Random Simulations and Results

The random simulations are conducted on randomly generated DDG with datasets of random sizes, generation times and usage frequencies. In the experiments, we randomly generate large DDGs with different number of datasets, each with a random size from 1GB to 100GB. The generation time is also random, from 10 hours to 100 hours. The usage frequency is again random, from once per month to once per year. In order to run our strategy, we partition the large DDGs into linear DDG segments with 50 datasets[12], on which we apply the T-CSB algorithm.

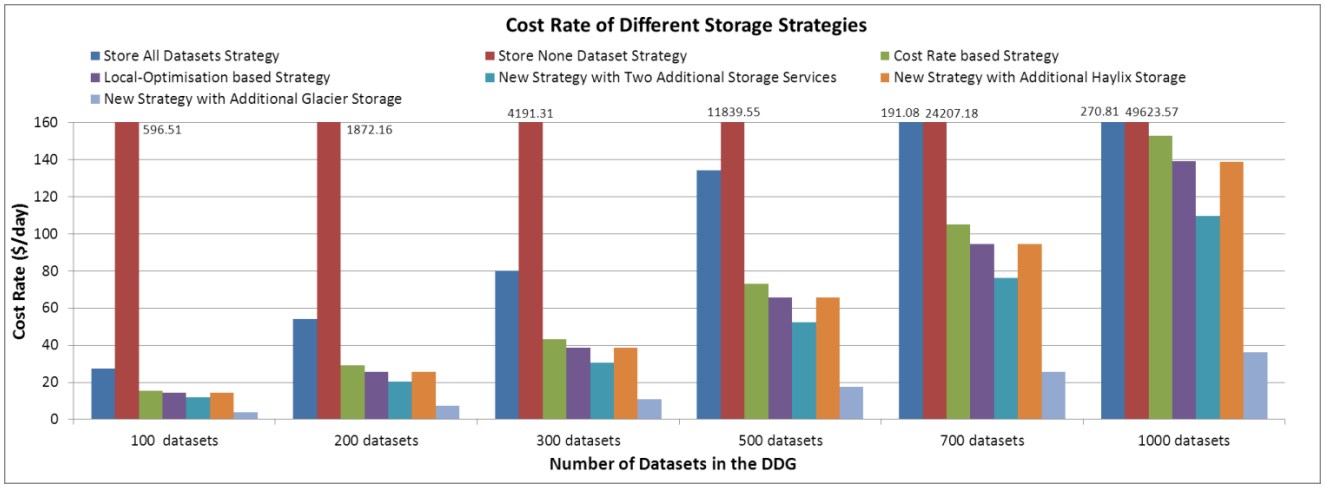

Figure 6. Cost effectiveness comparison of different storage strategies

TABLE I. DETAILED DATASETS STORAGE STATUS OF DIFFERENT STORAGE STRATEGIES

| Strategies / DDGs | Cost Rate based Strategy | | Local-Optimisation based Strategy | | New Strategy with Two Additional Storage Services | | | | New Strategy with Additional Haylix Storage | | | New Strategy with Additional Glacier Storage | | |
|---|---|---|---|---|---|---|---|---|---|---|---|---|---|---|
| | Deleted | Stored (S3) | Deleted | Stored (S3) | Deleted | Stored (S3) | Stored (Service1) | Stored (Service2) | Deleted | Stored (S3) | Stored (Haylix) | Deleted | Stored (S3) | Stored (Glacier) |
| 100 datasets | 64 | 36 | 57 | 43 | 43 | 0 | 29 | 28 | 57 | 38 | 5 | 5 | 0 | 95 |
| 200 datasets | 133 | 67 | 118 | 82 | 93 | 0 | 38 | 69 | 118 | 72 | 10 | 29 | 0 | 171 |
| 300 datasets | 203 | 97 | 176 | 124 | 149 | 0 | 69 | 82 | 173 | 110 | 17 | 29 | 0 | 271 |
| 500 datasets | 334 | 166 | 286 | 214 | 223 | 0 | 98 | 179 | 286 | 187 | 27 | 50 | 0 | 450 |
| 700 datasets | 466 | 234 | 406 | 294 | 324 | 0 | 150 | 226 | 404 | 262 | 34 | 67 | 0 | 633 |
| 1000 datasets | 644 | 356 | 577 | 423 | 428 | 0 | 182 | 390 | 573 | 379 | 48 | 103 | 0 | 897 |

*1) Cost Effectiveness Evaluation*

Based on the above settings, we run evaluation strategies on DDGs with different number of datasets and calculate the cost rates (i.e. average daily cost) of storing the datasets. Figure 6 shows the increases of the daily cost of different strategies as the number of datasets grows in the DDG, and Table I illustrates detailed datasets storage status of the DDGs under different storage strategies.

---

[12] The impact of DDG partition on cost effectiveness and efficiency of the strategy has been investigated in our prior work [36]     D. Yuan, Y. Yang, X. Liu, W. Li, L. Cui, M. Xu, *et al.*, "A Highly Practical Approach towards Achieving Minimum Datasets Storage Cost in the Cloud," *IEEE Transactions on Parallel and Distributed Systems,* vol. 24, pp. 1234-1244, 2012..

From Figure 6, we can see that the "store none dataset" and "store all datasets" strategies are very cost ineffective. By investigating the trade-off between computation and storage, the "cost rate based strategy" and "local-optimisation based strategy" can smartly choose to store or delete the datasets in one cloud storage service (as shown in Table I), thereby largely reducing the cost rate for storing datasets with one cloud service provider. If more cloud storage services are available, as shown in Figure 6, the simulation of "new strategy with two additional storage services" demonstrates further reduction of the cost rate by taking bandwidth cost into account. Table I shows the number of datasets transferred and smartly stored in two representative cloud storage services with our new strategy. Furthermore, how much cost can be reduced depends on the price of available storage services. In the simulation of "new strategy with additional Haylix storage", although some datasets are transferred to Haylix for storage (as shown in Table I), the cost rate only drops slightly comparing to the "local-optimisation based strategy" (as shown in Figure 6). This is because the price of Haylix is not much cheaper than Amazon S3 cloud. In contrast, in the simulation of "new strategy with additional Glacier storage", our new strategy significantly reduces the cost rate (as shown in Figure 6) by transferring datasets to Glacier[13] for storage (as shown in Table I).

From the above simulation, we can see that for different price models of cloud storage services, our new strategy can always store the datasets accordingly, even in the situation that the price difference is minor (e.g. the simulation of "new strategy with additional Haylix storage"). Hence our strategy is very effective in reducing the cost (i.e. cost-effective).

*2) Efficiency Evaluation*

In this sub-section, we conduct two sets of experiments to evaluate the efficiency of our storage strategy, where the experimental results are illustrated in Figure 7.

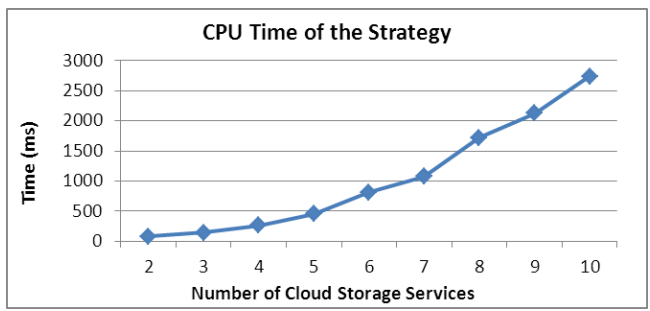

(a) A 100 datasets DDG with different number of cloud storage services.

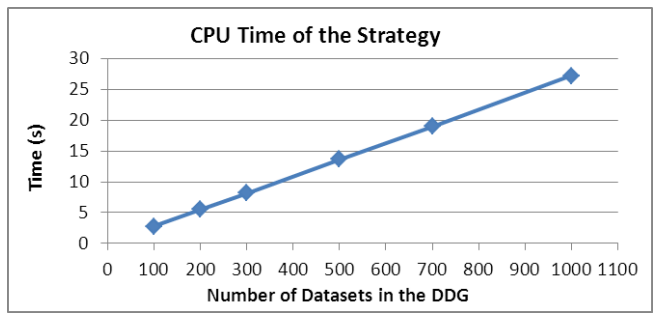

(b) Different DDGs with 10 cloud storage services

Figure 7. Efficiency evaluation of our strategy

---

[13] Data stored in Glacier usually need 3 to 5 hours to become available when users retrieve them. As analysed in Section 2.2, users' delay tolerance is out of the scope of this paper. Hence we only focus on the cost in the simulation.

In Figure 7 (a), we run our strategy on a 100 datasets DDG with different numbers of cloud storage services. As we can see, even with 10 storage services, the computation of our strategy finishes in less than 3 seconds, which is very efficient. Next, in Figure 7 (b), we run our strategy with 10 storage services on DDGs with different numbers of datasets. As the increase of datasets number in DDGs, we can see a linear growth of CPU time of the strategy, and even for the DDG with 1000 datasets, our strategy can finish within 30 seconds. This is because we have adapted the philosophy of local-optimisation in our strategy and partitioned the large DDG into segments with 50 datasets (i.e., $n_i=50$, $l=1, 2, \ldots 20$). These experimental results are also consistent with the time complexity of our strategy (i.e., $O(lm^2n_i^4)$) presented at the end of Section 4.3. Hence, we deem that our strategy is very efficient for runtime utilisation in the cloud.

## 5.3 Case Studies

The general random simulations demonstrate the general performance of our datasets storage strategy. In this sub-section, we investigate three real scientific applications as case studies to show how our storage strategy works in the real world.

The applications are 1) the FEM application introduced in Section 2, 2) a Climatological Analyses Application in Meteorology and 3) a Pulsar Searching Application in Astrophysics. From the domain scientists, we get the DDGs of these applications as well as real statistics of the datasets. If these applications are migrated to the cloud, we assume that the applications be deployed in Amazon EC2 cloud with S3 storage service. Furthermore, we assume the generated datasets can be transferred to Haylix or Glacier for storage. Based on these settings, we apply the evaluation strategies introduced in Section 5.1 to the applications and demonstrate the results.

### 1) Finite Element Modelling Application

In the FEM application, one execution of the workflow generates four datasets. Scientists may need to re-analyse these datasets, or reuse them in new workflows and generate new datasets. As the application runs, the workflow needs to be executed very often; therefore a large number of datasets are generated. Figure 8 shows a DDG segment generated in the execution of the FEM workflow. The sizes and generation times of datasets are from Swinburne Structural Mechanics research group of running this application in local computing facilities. The usage frequencies of datasets are also based on researchers' understanding of the application.

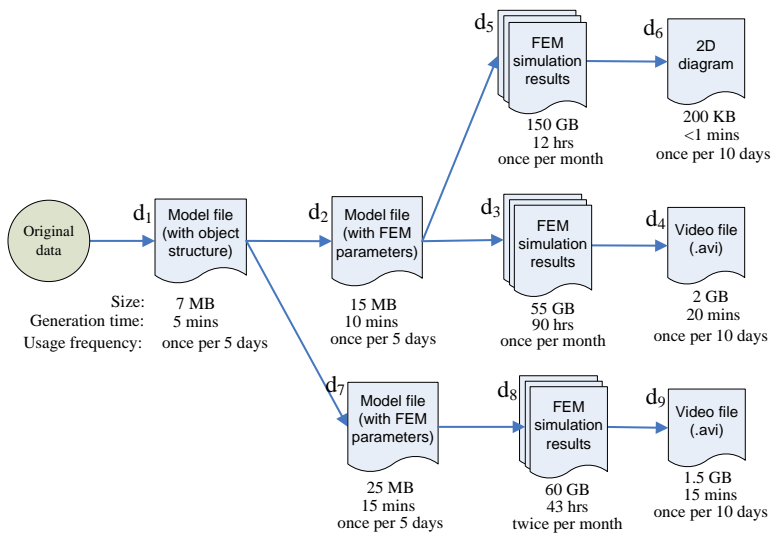

Figure 8. A DDG segment of the FEM application

Table II demonstrates datasets' storage status under different strategies and the corresponding monthly cost of the DDG segment. Furthermore, we estimate the yearly cost of storing all generated datasets by the FEM application from the Structural Mechanics Research Group and demonstrate the results in Table II (the far right column) as well.

TABLE II. DATASETS STORAGE STATUS AND COST OF THE FEM APPLICATION

| Strategies \ Datasets | Model files | | | FEM simulation results | | | Video files | | 2D diagram | Monthly cost of the DDG segment | Total cost of all datasets in one year |
|---|---|---|---|---|---|---|---|---|---|---|---|
| | $d_1$ | $d_2$ | $d_7$ | $d_3$ | $d_5$ | $d_8$ | $d_4$ | $d_9$ | $d_6$ | | |
| 1) Store all datasets strategy | Stored (S3) | Stored (S3) | Stored (S3) | Stored (S3) | Stored (S3) | Stored (S3) | Stored (S3) | Stored (S3) | Stored (S3) | $40.12 | $10130 |
| 2) Store none datasets strategy | Deleted | Deleted | Deleted | Deleted | Deleted | Deleted | Deleted | Deleted | Deleted | $58.3 | $14575 |
| 3) Cost rate based strategy | Stored (S3) | Stored (S3) | Stored (S3) | Stored (S3) | Deleted | Deleted | Deleted | Stored (S3) | Stored (S3) | $18.8 | $4800 |
| 4) Local-optimisation based strategy | Stored (S3) | Stored (S3) | Stored (S3) | Stored (S3) | Deleted | Stored (S3) | Deleted | Deleted | Stored (S3) | $18.6 | $4670 |
| 5) New strategy with additional Haylix storage | Stored (S3) | Stored (S3) | Stored (S3) | Stored (S3) | Deleted | Stored (S3) | Deleted | Deleted | Stored (S3) | $18.6 | $4670 |
| 6) New strategy with additional Glacier storage | Stored (Glacier) | Stored (Glacier) | Stored (Glacier) | Stored (Glacier) | Deleted | Stored (Glacier) | Stored (Glacier) | Stored (Glacier) | Stored (S3) | $3.32 | $850 |

*2) Climatological Analyses Application*

In this case study, we present the Global Positioning System (GPS) Radio Occultation (RO) data retrieval and climatological analyses processes, which is currently undertaken at the RMIT SPACE research centre.

GPS signals received by GPS receivers on-board Earth Observation satellites have been approved as a useful data source for retrieving Earth's atmospheric profiles (such as temperature, pressure and water vapours). To obtain high accuracy and high resolution of the atmospheric profiles, large volume of satellite and meteorological data are required in the complex data retrieval processes. There are two main stages in the whole processing. The first stage is the atmospheric profile retrieval stage which processes the GPS signals, with other inputs (atmospheric models and geometry data), to retrieve atmospheric profiles. The second stage is the meteorological analyses stage where three different studies are preformed and three output data products are generated.

Researchers conduct these climatological analyses in different areas globally. Figure 9 is a DDG segment generated by the RO retrieval and climatological analyses processes. The input data are accumulated GPS signal data for 10 years. The size of the data involved and process time at each stage are from an example that focuses on the Australian regions which are the main interested of Australian meteorologists. From the researchers' understanding these data are normally reused twice per month.

Table III demonstrates datasets' storage status under different strategies and the corresponding monthly cost of the DDG segment. Furthermore, we estimate the yearly cost of storing all generated datasets by the Climatological Analyses Application for analysing global GPS signal data and demonstrate the results in Table III (the far right column) as well.

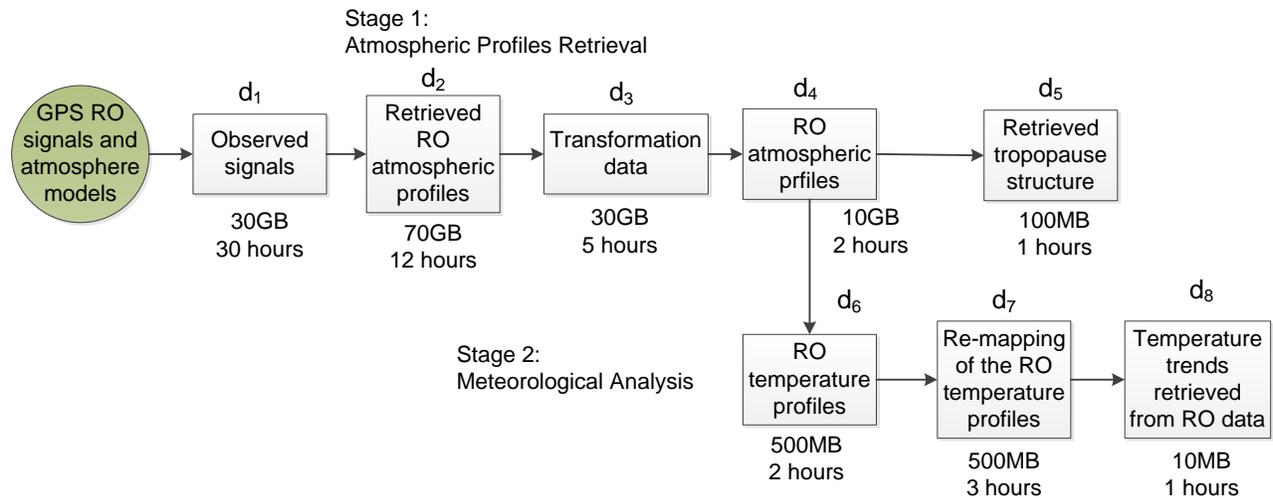

**Figure 9. DDG of Climatological Analyses Application**

TABLE III. DATASETS STORAGE STATUS AND COST OF THE CLIMATOLOGICAL ANALYSES APPLICATION

| Strategies \ Datasets | Stage 1: Atmospheric profiles retrieval | | | | | Stage 2: Meteorological analysis | | | Monthly cost of the DDG segment | Total cost of global datasets in one year |
|---|---|---|---|---|---|---|---|---|---|---|
| | $d_1$ | $d_2$ | $d_3$ | $d_4$ | $d_5$ | $d_6$ | $d_7$ | $d_8$ | | |
| 1) Store all datasets strategy | Stored (S3) | Stored (S3) | Stored (S3) | Stored (S3) | Stored (S3) | Stored (S3) | Stored (S3) | Stored (S3) | $21.17 | $16845.85 |
| 2) Store none datasets strategy | Deleted | Deleted | Deleted | Deleted | Deleted | Deleted | Deleted | Deleted | $75.6 | $60158.07 |
| 3) Cost rate based strategy | Stored (S3) | Deleted | Deleted | Stored (S3) | Stored (S3) | Stored (S3) | Stored (S3) | Stored (S3) | $11.97 | $9525.03 |
| 4) Local-optimisation based strategy | Stored (S3) | Deleted | Deleted | Stored (S3) | Stored (S3) | Stored (S3) | Stored (S3) | Stored (S3) | $11.97 | $9525.03 |
| 5) New strategy with additional Haylix storage | Stored (S3) | Deleted | Deleted | Stored (S3) | Stored (S3) | Stored (S3) | Stored (S3) | Stored (S3) | $11.97 | $9525.03 |
| 6) New strategy with additional Glacier storage | Stored (Glacier) | Stored (Glacier) | Stored (Glacier) | Stored (Glacier) | Stored (Glacier) | Stored (Glacier) | Stored (Glacier) | Stored (Glacier) | $7.06 | $5617.94 |

*3) Pulsar Searching Application*

Swinburne Astrophysics group has been conducting pulsar searching surveys using the observation data from Parkes Radio Telescope, which is one of the most famous radio telescopes in the world[14]. Pulsar searching is a typical scientific application that contains complex and time consuming tasks and needs to process terabytes of data.

The execution of the application has two main stages: Files Preparation and Seeking Candidates, where in each phase three datasets are generated as shown in Figure 10. For illustration, a DDG segment generated in this application for processing **one** hour's observation data is shown in Figure 10, as well as the sizes and generation times which are from running this application on Swinburne Astrophysics Supercomputer. From Swinburne Astrophysics research group, we understand that the "De-dispersion files" is the most useful dataset. Based on these files, many accelerating and seeking methods can be used to search pulsar candidates. Based on the scenario, we set the "De-dispersion files" to be used once every 4 days and other datasets to be used once every 10 days.

---

[14] http://www.parkes.atnf.csiro.au/

Table IV demonstrates datasets' storage status under different strategies and the corresponding monthly cost of the DDG segment. Furthermore, we estimate the yearly cost of storing all generated datasets by the Pulsar Searching Application for analysing a whole day's observation data and demonstrate the results in Table IV (the far right column) as well.

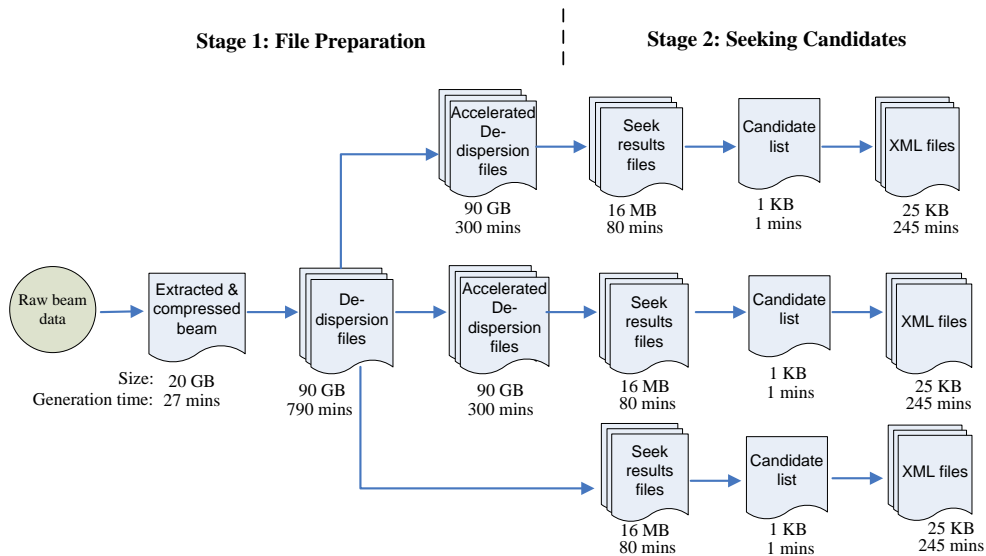

Figure 10. DDG of pulsar searching application

TABLE IV. DATASETS STORAGE STATUS AND COST OF THE PULSAR SEARCHING APPLICATION

| Strategies \ Datasets | Extracted beam | De-dispersion files | Accelerated de-dispersion files | Seek results | Pulsar candidates | XML files | Monthly cost of the DDG segment | Total cost of datasets in one year |
|---|---|---|---|---|---|---|---|---|
| 1) Store all datasets strategy | Stored (S3) | Stored (S3) | Stored (S3) | Stored (S3) | Stored (S3) | Stored (S3) | $43.5 | $12528 |
| 2) Store none datasets strategy | Deleted | Deleted | Deleted | Deleted | Deleted | Deleted | $73.9 | $21283.2 |
| 3) Cost rate based strategy | Deleted | Stored (deleted initially) | Deleted | Stored (S3) | Deleted | Stored (S3) | $17.1 | $4924.8 |
| 4) Local-optimisation based strategy | Deleted | Stored (S3) | Deleted | Stored (S3) | Deleted | Stored (S3) | $16.65 | $4795.2 |
| 5) New strategy with additional Haylix storage | Deleted | Stored (S3) | Deleted | Stored (S3) | Deleted | Stored (S3) | $16.65 | $4795.2 |
| 6) New strategy with additional Glacier storage | Deleted | Stored (S3) | Deleted | Stored (S3) | Deleted | Stored (Glacier) | $16.65 | $4795.2 |

*4) Summary*

From the three case studies we can see that, by only considering the trade-off between computation and storage, the "cost rate based strategy" and "local-optimisation based strategy" greatly reduce the monthly cost comparing to the "store all datasets strategy" and "store none datasets strategy". When considering Haylix as the cheaper cloud storage for data transfer, our new strategy turns out to be the same as the "local-optimisation based strategy". This is because the Haylix storage cloud does not have much advantage in price comparing to Amazon S3. In contrast, when we consider Glacier as the cheaper cloud storage for data transfer, our new strategy performs differently in the case studies. 1) In the FEM application, most of the datasets in the DDG segment are transferred to Glacier for storage and further reduces 82.15% of the monthly cost comparing to the "local-optimisation based strategy". 2) In the Climatological Analyses application, all the datasets in the DDG segment are transferred to Glaciers for storage and further reduces 41.02% of the monthly cost comparing to the "local-optimisation based strategy". 3) However, in the Pulsar Searching application, only one dataset is transferred to Glacier for storage and the monthly cost also does not change noticeably. This is because Glacier is not a cost-effective option for transferring data according to the features of the Pulsar Searching application. Hence we can draw the conclusion that our new storage strategy is cost effective and generic to be utilised in different applications.

## 6 RELATED WORK

Today, research on scientific applications in the cloud becomes popular [17] [18] [25] [30]. Comparing to the traditional computing systems, e.g. cluster, grid and HPC systems, a cloud computing system has cost benefits in various aspects [4]. With Amazon clouds' cost model and BOINC volunteer computing middleware, the work in [19] analyses the cost benefits of cloud computing versus grid computing. The work by Deelman et al. [11] also applies Amazon clouds' cost model and demonstrates that cloud computing offers a cost-effective way to deploy scientific applications. The work mentioned above mainly focuses on the comparison of cloud computing systems and the traditional distributed computing paradigms, which shows that applications running in the cloud have cost benefits. However, our work focuses on reducing cost for running application in the cloud.

This paper is mainly inspired by the research in the area of scheduling, in which much work focuses on reducing various "costs" for applications [29], systems [31] or data centre networks [10]. The difference is that scheduling aims at improving resource utilisation whilst our work investigates the trade-off among computation, storage and bandwidth costs, which is a unique issue in cloud computing due to the pay-as-you-go model. Another important foundation for our work is the research on data provenance. Due to the importance of data provenance in scientific applications, many works about recording data provenance of the system have been done [9]. Recently, research on data provenance in cloud computing systems has also appeared [22]. More specifically, Osterweil et al. [24] present how to generate a data derivation graph for execution of a scientific workflow. Foster et al. [12] propose the concept of virtual data in the Chimera system, which enables the automatic regeneration of datasets when needed. Our DDG is based on data provenance, which depicts the dependency relationships of all the generated datasets in the cloud. With DDG, we can manage where the datasets are stored or how to regenerate them.

As the trade-off among computation, storage and bandwidth is an important issue in the cloud, much research has already embarked on this issue to a certain extent. First, plenty of research has been done with regard to the trade-off between computation and storage. The Nectar system [15] is designed for automatic management of data and computation in data centres, where obsolete datasets are deleted and regenerated whenever reused in order to improve resource utilisation. In [11], Deelman et al. present that storing some popular intermediate data can save the cost in comparison to always regenerating them from the input data. In [2], Adams et al. propose a model to represent the trade-off of computation cost and storage cost. In [35], the authors propose the CTT-SP algorithm that can find the best trade-off between computation and storage in the cloud, based on which a highly cost-effective and practical strategy is developed for storing datasets with one cloud service provider [36]. However, the above work did not consider bandwidth cost into the trade-off model. In [6], Baliga et al. investigate the trade-off among computation, storage and bandwidth in the infrastructure level of cloud systems, where reducing energy consumption is the main research goal. In [3], Agarwala et al. transform application data to certain formats and store them with different cloud services in order to reduce storage cost in the cloud, but data dependency and the option of data regeneration are not considered in their work. In this paper, we propose the T-CSB algorithm which can find the best trade-off among computation, storage and bandwidth costs for storing datasets of linear DDG in the cloud. Based on this algorithm, we develop a cost-effective runtime strategy for storing generated application datasets with multiple service providers in the cloud.

## 7 CONCLUSIONS AND FUTURE WORK

In this paper, we have investigated the unique features of storing large volume of generated scientific datasets with multiple cloud service providers in the cloud. Towards achieving the cost effectiveness, we have proposed a T-CSB (Trade-off among Computation, Storage and Bandwidth) algorithm to find the minimum cost storage strategy for datasets of linear DDG, which also represents the best trade-off among three key factors (computation, storage and bandwidth) for the cost of data storage in the cloud. Based on the algorithm, we have developed a local-optimisation based datasets storage strategy for multiple service providers in the cloud. General random simulations and three specific case studies indicate that our strategy is very cost effective with highly practical runtime efficiency.

In our current work, we assume that the storage of one cloud service provider have a unified price. However, in the real world, the price of cloud storage is different according to different usages. In the future, we will incorporate more complex pricing models in our datasets storage cost model. Furthermore, methods for forecasting dataset usage frequency can be further studied, with which our storage strategy can be adapted to different types of applications more easily.

## ACKNOWLEDGMENT

The research work reported here is partly supported by Australian Research Council under DP110101340 and LP130100324. We are also grateful for discussions on the Finite Element Modelling application with Dr. S. Xu and on the Pulsar Searching application with Dr. W. van Straten and Dr. L. Levin, from Faculty of Science, Engineering and Technology, Swinburne University of Technology.

## REFERENCES

[1] *Amazon Cloud Services*. Available: http://aws.amazon.com/
[2] I. Adams, D. D. E. Long, E. L. Miller, S. Pasupathy, and M. W. Storer, "Maximizing Efficiency by Trading Storage for Computation," in *Workshop on Hot Topics in Cloud Computing (HotCloud'09)*, San Diego, CA, 2009, pp. 1-5.
[3] S. Agarwala, D. Jadav, and L. A. Bathen, "iCostale: Adaptive Cost Optimization for Storage Clouds," in *IEEE International Conference on Cloud Computing (CLOUD2011)*, 2011, pp. 436-443.
[4] M. Armbrust, A. Fox, R. Griffith, A. D. Joseph, R. Katz, A. Konwinski*, et al.*, "A View of Cloud Computing," *Communication of the ACM,* vol. 53, pp. 50-58, 2010.
[5] M. S. Avila-Garcia, X. Xiong, A. E. Trefethen, C. Crichton, A. Tsui, and P. Hu, "A Virtual Research Environment for Cancer Imaging Research," in *7th International Conference on E-Science (e-Science2011)*, 2011, pp. 1-6.
[6] J. Baliga, R. W. Ayre, K. Hinton, and R. S. Tucker, "Green cloud computing: Balancing energy in processing, storage, and transport," *Proceedings of the IEEE,* vol. 99, pp. 149-167, 2011.
[7] R. Bose and J. Frew, "Lineage Retrieval for Scientific Data Processing: A Survey," *ACM Computing Surveys,* vol. 37, pp. 1-28, 2005.
[8] A. Burton and A. Treloar, "Publish My Data: A Composition of Services from ANDS and ARCS," in *5th IEEE International Conference on e-Science (e-Science'09),*, Oxford, UK, 2009, pp. 164-170.
[9] P. Chen, B. Plale, and M. S. Aktas, "Temporal representation for scientific data provenance," in *8th International Conference on E-Science (e-Science2012)*, 2012, pp. 1-8.
[10] Y. Cui, H. Wang, and X. Cheng, "Channel Allocation in Wireless Data Center Networks," in *IEEE INFOCOM 2011* 2011, pp. 1395-1403.
[11] E. Deelman, G. Singh, M. Livny, B. Berriman, and J. Good, "The Cost of Doing Science on the Cloud: the Montage Example," in *ACM/IEEE Conference on Supercomputing (SC'08)*, Austin, Texas, 2008, pp. 1-12.
[12] I. Foster, J. Vockler, M. Wilde, and Z. Yong, "Chimera: A Virtual Data System for Representing, Querying, and Automating Data Derivation," in *14th International Conference on Scientific and Statistical Database Management, (SSDBM'02)*, Edinburgh, Scotland, UK, 2002, pp. 37-46.
[13] I. Foster, Z. Yong, I. Raicu, and S. Lu, "Cloud Computing and Grid Computing 360-Degree Compared," in *Grid Computing Environments Workshop (GCE'08)*, Austin, Texas, USA, 2008, pp. 1-10.
[14] S. K. Garg, R. Buyya, and H. J. Siegel, "Time and Cost Trade-Off Management for Scheduling Parallel Applications on Utility Grids," *Future Generation Computer Systems,* vol. 26, pp. 1344-1355, 2010.

[15] P. K. Gunda, L. Ravindranath, C. A. Thekkath, Y. Yu, and L. Zhuang, "Nectar: Automatic Management of Data and Computation in Datacenters," in *9th Symposium on Operating Systems Design and Implementation (OSDI'2010)*, Vancouver, BC, Canada, 2010, pp. 1-14.

[16] X. Huang, Z. Luo, and B. Yan, "Cyberinfrastructure and e-Science Application Practices in Chinese Academy of Sciences," in *7th International Conference on E-Science (e-Science2011)*, 2011, pp. 348-354.

[17] M. Humphrey, N. Beekwilder, J. L. Goodall, and M. B. Ercan, "Calibration of watershed models using cloud computing," in *8th International Conference on E-Science (e-Science2012)*, 2012, pp. 1-8.

[18] G. Juve, E. Deelman, K. Vahi, and G. Mehta, "Data Sharing Options for Scientific Workflows on Amazon EC2," in *ACM/IEEE Conference on Supercomputing (SC'10)*, New Orleans, Louisiana, USA, 2010, pp. 1-9.

[19] D. Kondo, B. Javadi, P. Malecot, F. Cappello, and D. P. Anderson, "Cost-Benefit Analysis of Cloud Computing versus Desktop Grids," in *23th IEEE International Parallel & Distributed Processing Symposium (IPDPS'09)*, Rome, Italy, 2009, pp. 1-12.

[20] X. Liu, Z. Ni, D. Yuan, Y. Jiang, Z. Wu, J. Chen, *et al.*, "A Novel Statistical Time-Series Pattern based Interval Forecasting Strategy for Activity Durations in Workflow Systems," *Journal of Systems and Software* vol. 84, pp. 354-376, 2011.

[21] B. Ludascher, I. Altintas, C. Berkley, D. Higgins, E. Jaeger, M. Jones, *et al.*, "Scientific Workflow Management and the Kepler System," *Concurrency and Computation: Practice and Experience,* pp. 1039–1065, 2005.

[22] K.-K. Muniswamy-Reddy, P. Macko, and M. Seltzer, "Provenance for the Cloud," in *8th USENIX Conference on File and Storage Technology (FAST'10)*, San Jose, CA, USA, 2010, pp. 197-210.

[23] H. Nguyen and D. Abramson, "WorkWays: Interactive Workflow-based Science Gateways," in *8th International Conference on E-Science (e-Science2012)*, 2012, pp. 1-8.

[24] L. J. Osterweil, L. A. Clarke, A. M. Ellison, R. Podorozhny, A. Wise, E. Boose, *et al.*, "Experience in Using A Process Language to Define Scientific Workflow and Generate Dataset Provenance," in *16th ACM SIGSOFT International Symposium on Foundations of Software Engineering*, Atlanta, Georgia, 2008, pp. 319-329.

[25] J. Qiu, J. Ekanayake, T. Gunarathne, J. Y. Choi, S.-H. Bae, H. Li, *et al.*, "Hybrid Cloud and Cluster Computing Paradigms for Life Science Applications " *Journal of BMC Bioinformatics,* vol. 11, 2010.

[26] X. Su, Y. Ma, H. Yang, X. Chang, K. Nan, J. Xu, *et al.*, "An Open-Source Collaboration Environment for Metagenomics Research," in *7th International Conference on E-Science (e-Science2011)*, 2011, pp. 7-14.

[27] A. S. Szalay and J. Gray, "Science in an Exponential World," *Nature,* vol. 440, pp. 23-24, 2006.

[28] S. Toor, M. Sabesan, S. Holmgren, and T. Risch, "A Scalable Architecture for e-Science Data Management," in *7th International Conference on E-Science (e-Science2011)*, 2011, pp. 210-217.

[29] D. Warneke and O. Kao, "Exploiting Dynamic Resource Allocation for Efficient Parallel Data Processing in the Cloud," *IEEE Transactions on Parallel and Distributed Systems,* vol. 22, pp. 985-997, 2011.

[30] Y. Yang, K. Liu, J. Chen, X. Liu, D. Yuan, and H. Jin, "An Algorithm in SwinDeW-C for Scheduling Transaction-Intensive Cost-Constrained Cloud Workflows," in *4th IEEE International Conference on E-Science (e-Science2008)*, 2008, pp. 374-375.

[31] L. Young Choon and A. Y. Zomaya, "Energy Conscious Scheduling for Distributed Computing Systems under Different Operating Conditions," *IEEE Transactions on Parallel and Distributed Systems,* vol. 22, pp. 1374-1381, 2011.

[32] D. Yuan, X. Liu, L. Cui, T. Zhang, W. Li, D. Cao and Y. Yang, "An Algorithm for Cost-Effectively Storing Scientific Datasets with Multiple Service Providers in the Cloud," in *9th International Conference on e-Science (e-Science2013),* , 2013, pp. 285-292.

[33] D. Yuan, Y. Yang, X. Liu, and J. Chen, "A Cost-Effective Strategy for Intermediate Data Storage in Scientific Cloud Workflows," in *24th IEEE International Parallel & Distributed Processing Symposium*, USA, 2010, pp. 1-12.

[34] D. Yuan, Y. Yang, X. Liu, and J. Chen, "A Local-Optimisation based Strategy for Cost-Effective Datasets Storage of Scientific Applications in the Cloud," in *IEEE International Conference on Cloud Computing*, Washington DC, USA, 2011, pp. 179-186.

[35] D. Yuan, Y. Yang, X. Liu, and J. Chen, "On-demand Minimum Cost Benchmarking for Intermediate Datasets Storage in Scientific Cloud Workflow Systems," *Journal of Parallel and Distributed Computing,* vol. 71, pp. 316-332, 2011.

[36] D. Yuan, Y. Yang, X. Liu, W. Li, L. Cui, M. Xu, *et al.*, "A Highly Practical Approach towards Achieving Minimum Datasets Storage Cost in the Cloud," *IEEE Transactions on Parallel and Distributed Systems,* vol. 24, pp. 1234-1244, 2012.

[37] D. Yuan, Y. Yang, X. Liu, G. Zhang, and J. Chen, "A Data Dependency Based Strategy for Intermediate Data Storage in Scientific Cloud Workflow Systems," *Concurrency and Computation: Practice and Experience,* vol. 24, pp. 956-976, 2010.

[38] M. Zaharia, A. Konwinski, A. D. Joseph, R. Katz, and I. Stoica, "Improving MapReduce Performance in Heterogeneous Environments," in *8th USENIX Symposium on Operating Systems Design and Implementation (OSDI'2008)*, San Diego, CA, USA, 2008, pp. 29-42.